{\catcode`\@=11
\gdef\SchlangeUnter#1#2{\lower2pt\vbox{\baselineskip 0pt \lineskip0pt
  \ialign{$\m@th#1\hfil##\hfil$\crcr#2\crcr\sim\crcr}}}
    }
    \def\gtrsim{\mathrel{\mathpalette\SchlangeUnter>}}
    \def\lesssim{\mathrel{\mathpalette\SchlangeUnter<}}
\documentstyle[editedvolume]{crckapb}
\begin{opening}
\title{Association of X-ray Quasars with Active Galaxies}
\author{H. ARP}   
\institute{Max-Planck-Institut f\"ur Astrophysik \\
	   Karl-Schwarzschild-Str.\ 1\\
	   85740 Garching, Germany}

\end{opening}

\runningtitle{Association of X-ray Quasars with Active Galaxies} 

\begin{document} 

\begin{abstract}
Analysis of ROSAT observations demonstrate that X-ray sources are associated
with bright Seyfert galaxies up to distances of about 40 arc min. These X-ray 
sources are predominantly identified with blue stellar objects (BSO's) some of which
are already catalogued as quasars.

The X-ray sources tend to pair and align across the nucleus of the active Seyfert. 
Enough redshifts are now available to indicate the similarities of the redshifts
in the quasar pairs and to enable computation of ejection velocities of about 0.1c.

BL Lac objects are a conspicuous kind of strong X-ray source which are associated
with Seyferts at a high level of probability. The BL Lac's appear to be a transition
form between quasars and compact companion galaxies. Both quasars and companion galaxies tend
to align along the minor axis of the ejecting galaxy and extend to the same maximum
separation of $\sim 400$ kpc. 

These observations require high redshift quasars to evolve into low redshift companion
galaxies. The initially high intrinsic redshift of the quasars must then arise from
the low particle masses in their relatively recently created matter.
\end{abstract}

\section{Introduction}
In 1966 it was discovered that radio quasars tended to pair across
active galaxies. Evidence that these quasars were, like other radio
sources, ejected from the energetic nuclei of the galaxies is reviewed
by Arp (1987). In the past decade X-ray telescopes efficiently
discovered point sources within fields of $\sim 1^{\rm o}$
radius. When these fields were centered on the especially strong X-ray
galaxies called Seyferts, it became obvious that there were physically
associated, excess numbers of point X-ray sources out to about
40$^\prime$ radius (Radecke 1997). These sources are overwhelmingly
quasars and confirmed very strongly their pairing and alignment across
the central, low redshift galaxies (Arp 1997a). An example is shown in
Fig.~1.

\unitlength1cm
\begin{picture}(0,9.5)
\put(1,0){\footnotesize } 
\end{picture}

\noindent{\footnotesize {\it Figure 1} The Sefyert 1 galaxy NGC~4235 with strong X-ray sources paired across it
(268 and 119
counts/ksec). Redshifts of z=.334 for the quasar and z=.136 for the BL
Lac object are much higher then the redshift of the central galaxy,
z=.007 (From Arp 1997b).}

\section{Empirical Characteristics of the Associations}

Fig.~2 incorporates in one schematic diagram the properties observed
over the last 30 years for quasars associated with low redshift
galaxies. Characteristically the quasars emerge close to the present
galaxy with high redshift and low luminosity. As they travel outward
they decrease their redshift and increase their luminosity. At maximum
separation from their parent they tend to be relatively strong X-ray
emitters and have redshift in the $.1 \lesssim z \lesssim .3$
range. This is the region where BL Lac objects (a rare, active kind of
quasar in which energetic continuum outbursts swamp the usual quasar
emission lines) are encountered. BL Lac objects themselves tend to
eject new objects and they also show the first signs of underlying
stellar population. They appear to be the transition phase between the
quasars and the compact, young galaxies. Finally an empirical sequence
of forms can be traced from the compact companions to increasingly
normal companions as their intrinsic redshifts continue to drop.

\unitlength1cm
\begin{picture}(3,11)
\put(1,0){\footnotesize }
\end{picture}

\noindent {\footnotesize {\it Figure 2}~ Schematic representation of
quasars and companion galaxies found associated with central galaxies
from 1966 to present. The progression of characteristic is empirical
but is also required by the variable particle mass theory of Narlikar
and Arp (1993).}

\section{A Single, Striking Example, NGC 3516.}
One of the particularly active Seyferts where associated X-ray BSO's
were identified by Radecke and Arp was observed spectroscopically by
Yaoquan Chu with the Beijing 2.2 meter telescope Fig.~3 shows the
electrifying results of his quasar confirmations (Chu et al 1997). It
can be readily seen how the quasars decrease in redshift as they
extend away from the central Seyfert. They are aligned within $\pm 20$
degrees of the central galaxy's minor axis (a result by itself which
has only $10^{-4}$ chance of being accidental).

Moreover the 5 quasars and the 6$^{\rm th}$ BL Lac-type object all
have redshifts very close to the quantized redshift values which
quasars on average exhibit (Arp et al 1990). The importance of this
result is that the peculiar velocity component of the redshift must be
relatively small and the major portion of the redshift cannot be a
recessional velocity.

\unitlength1cm
\begin{picture}(3,10.5)
\put(1,0){\footnotesize } 
\end{picture}

\noindent{\footnotesize {\it Figure 3}~ All bright X-ray objects around
the very active Seyfert galaxy NGC~3516 which have had their redshifts
measured by Chu et al (1997). Redshifts are written to the upper right
of each quasar and quasar-like object.}

\section{Evolution of Quasars into Companion Galaxies.}
As Fig.~2 shows, the empirical continuity of properties suggests very
strongly that the quasars turn into normal companion galaxies as they
age. 

Among these properties are:
\vskip-3mm
\begin{itemize}
\item[1)] Intrinsic redshift. The quasars have high intrinsic redshift
marking them as much younger than the parent galaxy. Companion
galaxies as a class have small excess redshifts (Arp 1994) in
consonance with their having reduced their intrinsic redshift as they
aged from quasars. But if they were not slightly younger than the
dominant galaxy in the group what could explain the systematic
redshifts of the companion galaxies?
\item[2)] Quantization of redshift. Quasar redshift are quantized in
large steps. Galaxy redshifts are quantized in small steps (Tifft
1976; Guthrie and Napier 1996). It is unlikely that the
quantization is caused by different effects in quasars and
galaxies. Therefore the continuous change of this intrinsic property
is strong evidence for an evolutionary process.
\end{itemize}

In addition to these arguments there is the direct evidence from their
location with respect to the ejecting galaxy. As Fig.~4 shows, the
QSO's are ejected within $\pm 20$ degrees of the minor axis whereas
companion galaxies are found preferentially within $\pm 35$ degrees of
the minor axis. Both distributions extend out to about 400 kpc (Arp
1998). As the companions age they apparently deviate somewhat from their
initially plunging orbits back to their nucleus of origin. But to find
both quasars and companions in this same unique volume of minor axis
space is the most direct proof possible of their common origin.

\unitlength1cm
\begin{picture}(3,7)
\put(1,0){\footnotesize }
\end{picture}

\noindent{\footnotesize {\it Figure 4}~ Schematic representation of
distribution of companion galaxies along minor axes of disk galaxies
($\pm 35$ degrees from Holmberg 1969; Sulentic et al. 1978; Zaritsky
et al. 1997). Quasars are observed $\pm 20$ degrees from minor axis
(Arp 1998).}

\section{A Word about Theory.}
The only explanation for intrinsic redshifts for stars and galaxies is
that proposed by Narlikar and Arp (1993) where newly created matter
has particle masses $m\gtrsim 0$ which grow with time. Since the newly
created matter is in the form of energy it starts off travelling with the signal
velocity, $c$. As particle masses increase with time the translational
velocity slows to conserve momentum. This agrees with observation in
that the small radio knots emerging from the innermost regions of
galaxy nuclei are observed to travel with very nearly $c$. They could
only represent the proto quasars which will subsequently evolve into
quasars and then into companion galaxies. Calculations by Narlikar and
Das (1980) with this theory yielded maximum excursions of about 400
kpc as are now observed. 

The pairs of X-ray quasars now confirm results originally obtained 30
years ago, namely that by the time quasars have evolved to redshifts
of $z\sim 1$ they are moving outward with about $0.1 c$ (Arp 1968,
1996). As they move further out and evolve into companion galaxies
their translational velocities must drop to less than about $\pm 20$
km/sec in order not to wash out their observed redshift quantizations
of 37.5 and 72 km/sec.

The newly created plasma which is expelled from the ejecting nuclei
must be composed of low mass, high cross section particles. This is
closely analogous to the superfluid which Viktor Ambarzumian (1958)
proposed to be responsible for the formation of new galaxies. He came
to this conclusion by simply looking at photographs of galaxies which
he judged to be showing formation of new galaxies by ejection.

The peculiar velocities of the particles in the plasma must also slow
as they gain mass. That would produce cooling. So the plasma would
cool and gain mass -- certainly in the direction of forming self
gravitating systems. After 40 years it seems likely that we will have
to go back and recalculate all the analysis of ejected plasmoids,
shock waves, magnetic fields, polarizations, etc., that have been
made in connection with the famous active objects ejecting material in
jets. The aim would be to see if this new kind of plasma naturally
evolves into quasars and then into companion galaxies as empirically observed.

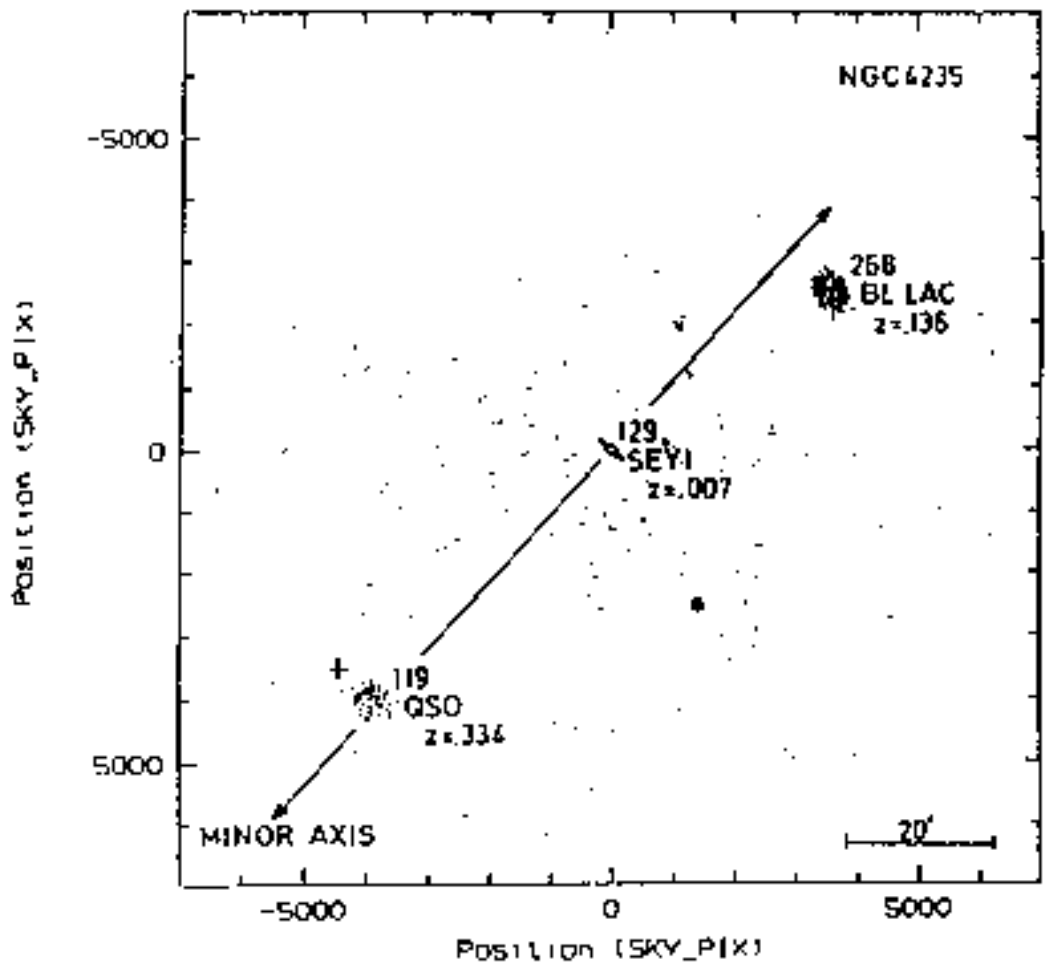
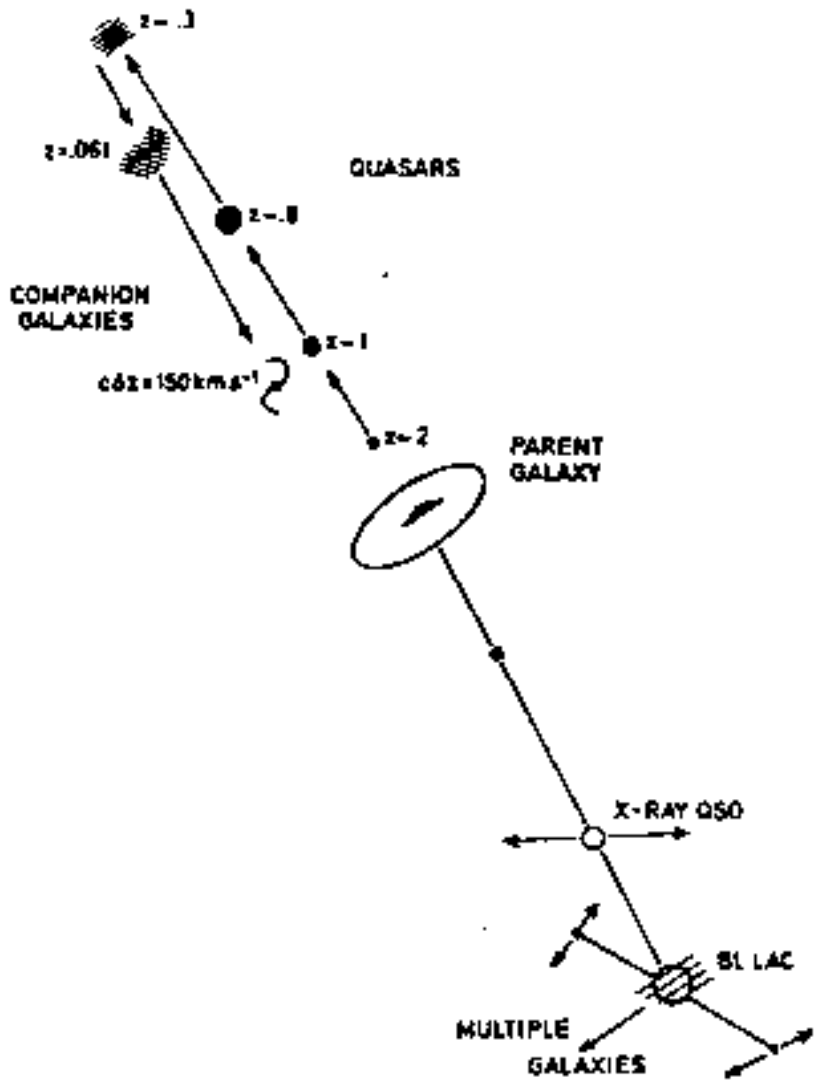
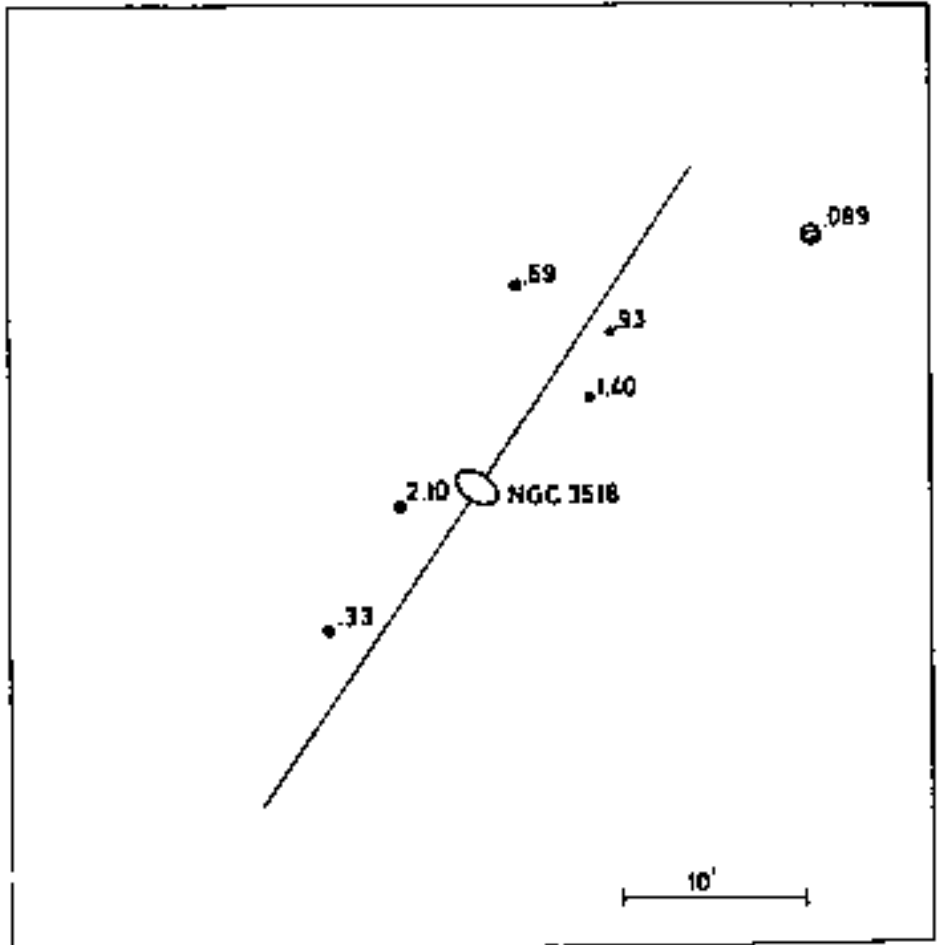
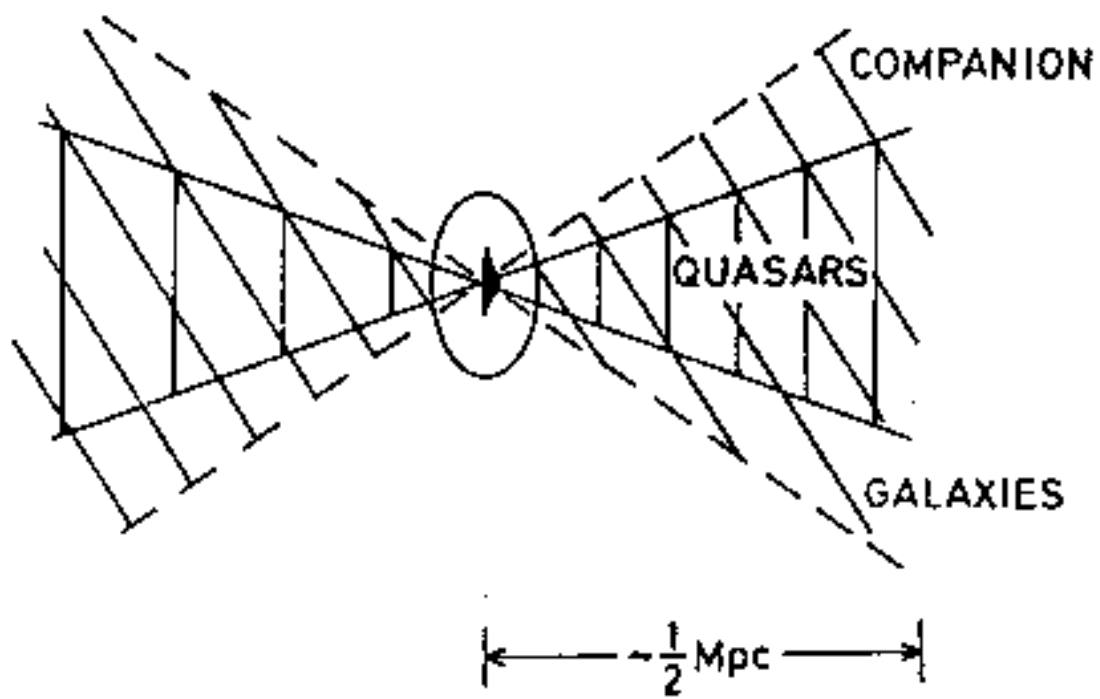
\end{document}